\documentclass[5p]{elsarticle}

\usepackage{hyperref}
\journal{Journal of \LaTeX\ Templates}

\begin{document}

\begin{frontmatter}

\title{Injection locking of a high power ultraviolet laser diode for laser cooling of ytterbium atoms}

\author{Toshiyuki Hosoya}
\author{Martin Miranda}
\author{Ryotaro Inoue} 
\author{Mikio Kozuma\corref{mycorrespondingauthor}}
\cortext[mycorrespondingauthor]{Corresponding author. Tel./fax: +81 3 5734 2451.}
\ead{kozuma@ap.titech.ac.jp}
\address{Department of Physics, Tokyo Institute of Technology, 2-12-1 O-Okayama, Meguro-ku, Tokyo 152-8550, Japan}
\begin{abstract}
We developed a high-power laser system at a wavelength of $399\ \mathrm{nm}$ for laser cooling of ytterbium atoms with ultraviolet laser diodes. The system is composed of an external cavity laser diode providing frequency stabilized output at a power of $40\ \mathrm{mW}$ and another laser diode for amplifying the laser power up to $220\ \mathrm{mW}$ by injection locking. The systematic method for optimization of our injection locking can also be applied to high power light sources at any other wavelengths. Our system, which does not depend on complex nonlinear frequency-doubling, has great importance for implementing transportable optical lattice clocks, and is also useful for investigations on condensed matter physics or quantum information processing using cold atoms.
\end{abstract}

\end{frontmatter}

% %\linenumbers
%
\section{Introduction}
\label{sec1}
Ytterbium (Yb) has five bosonic (${}^{168,170,172,174,176}$Yb) and two fermionic  (${}^{171,173}$Yb) isotopes with relatively high natural abundances, which enables us to perform various experiments on Bose-Bose and Fermi-Bose mixtures \cite{PhysRevA.79.021601,nphys2028}. Narrow intercombination transitions originated in two valence electrons are used not only for implementing frequency standards such as optical lattice clocks \cite{nature03541,Hinkley13092013} but also for probing phase transitions of quantum degenerate gases \cite{PhysRevLett.110.173201}. The ground-state electron configuration is in singlet state and is thus insensitive to environmental magnetic field, which is another advantage for the preparation of optical standards. Note that one of the isotopes, ${}^{171}$Yb, has a nuclear spin of 1/2, which is ideal for implementing a long-lived qubit \cite{s00340-009-3696-4,PhysRevA.81.042331,PhysRevA.81.062308,PhysRevA.84.030301,PhysRevLett.106.160501,PhysRevLett.110.163602}.

Experiments with the cold Yb atoms require high power light sources at a wavelength of $399\ \mathrm{nm}$ which are utilized for a variety of conventional techniques such as Zeeman slowing, magneto-optical trapping \cite{PhysRevA.59.R934,:/content/aip/journal/rsi/84/4/10.1063/1.4802682} and transverse two-dimensional cooling \cite{s11467-009-0033-7}. Unfortunately, the output power of ultraviolet laser diode systems has so far been limited to $80\ \mathrm{mW}$ \cite{1347-4065-42-8R-5059,1347-4065-42-7A-L754,doi:10.1117/12.733047,KAIST}. Frequency doubling using non-linear crystals such as LBO or BBO has conventionally been utilized for preparing high power light sources \cite{Adams199289,Pizzocaro:14}. However, they require locking of high finesse optical cavities, which results in complicated systems. While periodically poled structures, e.g., PPLN, PPKTP or their waveguides \cite{Tanimura:06,PhysRevLett.92.203602} can also be utilized to obtain high nonlinear conversion efficiency, they suffer from instability of the output power caused by unwanted optical losses such as photorefractive damage \cite{Volk:94,Villarroel:10} or gray track \cite{:/content/aip/journal/apl/65/19/10.1063/1.112688}.

In this manuscript, we develop a high-power laser system which is composed of an external cavity laser diode (ECLD) providing frequency stabilized output and another laser diode for amplifying the laser power by injection locking. The method employed here could be applicable to prepare high power light sources at other wavelengths such as $397\ \mathrm{nm}$ (Ca${}^{+}$ \cite{Urabe:92}), $401\ \mathrm{nm}$ (Er \cite{PhysRevLett.96.143005}), and $411\ \mathrm{nm}$ (Tm \cite{PhysRevA.82.011405} and Ho \cite{PhysRevA.89.041401}).
\section{Experiment Setup}
\label{sec2}
\begin{figure}[htbp]
 \begin{center}
  \includegraphics{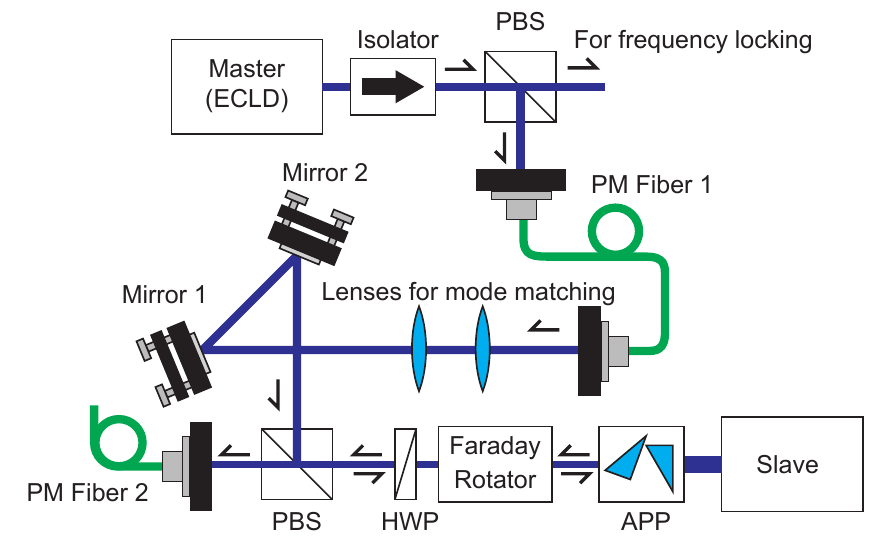}
 \caption{Schematic of the experimental setup. 
 PBS, polarization beam splitter;  PM Fiber, polarization maintained fiber;  HWP, half-wave plate;  APP, anamorphic prism pair;  ECLD, external cavity laser diode.
 }
 \label{fig1}
 \end{center}
\end{figure}

Figure \ref{fig1} shows the schematic of our experimental setup. An ECLD in a Littrow configuration is utilized as a master laser, where the external cavity length is set to $10\ \mathrm{mm}$. We use a commercial laser diode chip (Nichia NDV4B16) with a nominal CW output power of $300\ \mathrm{mW}$ and a central wavelength of $402\ \mathrm{nm}$ at a temperature of $25{}^\circ\mathrm{C}$. The focal length of the collimation lens (LightPath 357775) is $4.02\ \mathrm{mm}$ and a blazed holographic grating with 2400 grooves/mm is utilized. An output power of $40\ \mathrm{mW}$ is obtained at the desired wavelength ($399\ \mathrm{nm}$), where the injection current and the chip temperature are set to $80\ \mathrm{mA}$ and $18{}^\circ\mathrm{C}$, respectively. The frequency of the ECLD is locked by using a Doppler-free saturated absorption signal of the ${}^{1}$S${}_{0}$-${}^{1}$P${}_{1}$ transition in ${}^{174}$Yb atoms obtained with an Yb hollow-cathode lamp (Hamamatsu L2783-70NE-Yb) \cite{1347-4065-47-12R-8856}.

In order to amplify the laser power by injection locking, a portion of the ECLD output is coupled into a polarization maintained fiber (PM fiber 1), and injected into the slave laser after passing through two lenses and two mirrors (Mirror 1, 2) for spatial mode matching. The injected power (seed power) is set to $5\ \mathrm{mW}$ in all experiments reported here.  The power is required for long-term (a couple of hours, typically) stability of our injection locking. The slave laser is composed of a laser diode identical to that in the master laser and a collimation lens (Melles Griot 06GLC201) with a focal length of $6.5\ \mathrm{mm}$. The output wavelength of the slave laser is adjusted to nearby the desired wavelength by decreasing the chip temperature to $-16{}^\circ\mathrm{C}$. In order to prevent condensation, the housing of the laser diode chip is purged with dry nitrogen.
\section{Optimization of the injection locking}
\label{seac3}
In order to achieve stable and efficient injection locking, the spatial modes of the seed beam from the master ECLD and the slave one should be well matched. The spatial mode matching can be roughly achieved by coupling the seed beam, which is reflected from the facet of the slave laser, to an additional fiber (PM fiber 2), where we have coupled the slave laser output to the PM fiber 2 with approximately $60\%$ of coupling efficiency. The coupling efficiency includes other intrinsic optical losses such as the transmittance of the coupling lens and the fiber insertion loss. The output of the PM fiber 2 is fed into a confocal scanning Fabry-Perot cavity and the transmitted signal is monitored, where the free spectral range (FSR) and finesse of the cavity are $1.5\ \mathrm{GHz}$ and 200, respectively. 
When the frequency stabilized seed light is coupled to the Fabry-Perot cavity, a single peak is monitored within one FSR. On the other hand, no peaks are observed in the case of the free-running slave laser because of its broad spectrum. When the seed light is injected to the slave laser under the roughly mode-matched condition described above, multiple peaks typically appear in the cavity transmission within one FSR.

\begin{figure}[htbp]
 \begin{center}
  \includegraphics{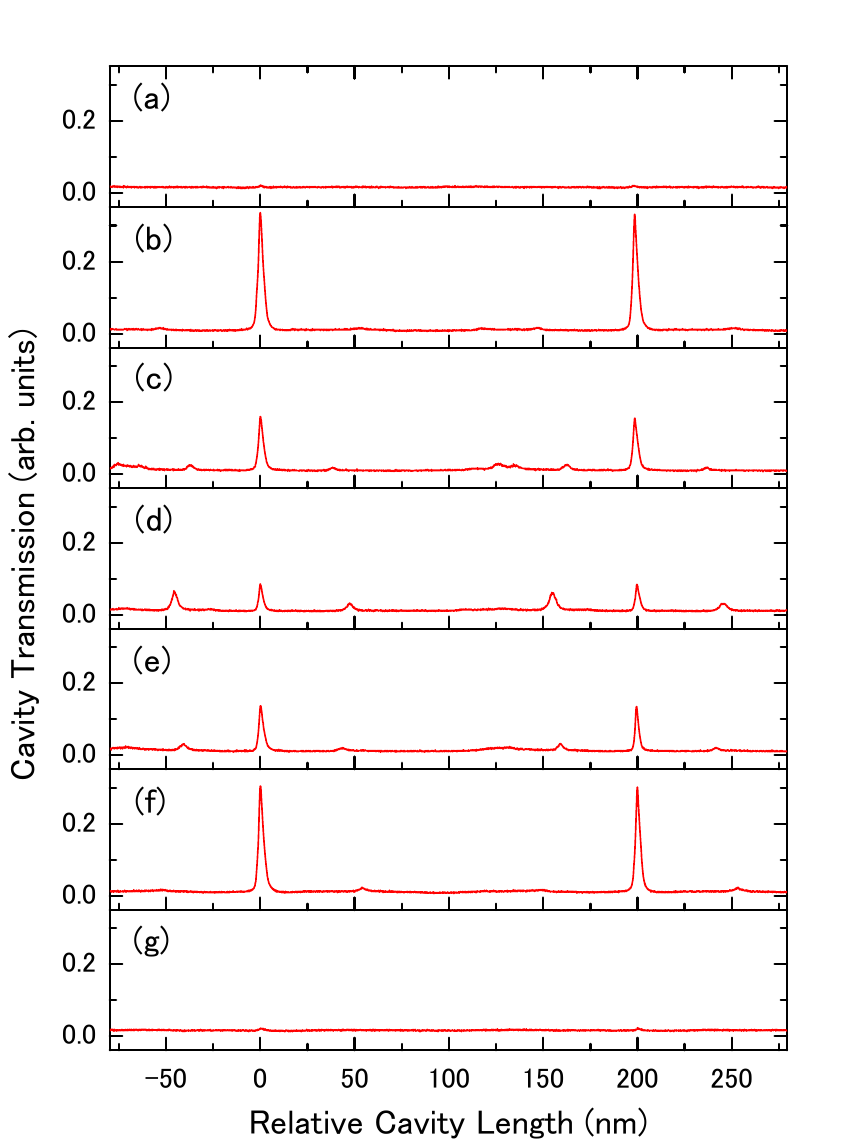}
 \caption{Variation of the scanning Fabry-Perot signals as the vertical angle of the Mirror 2 is gradually changed (increasing alphabetical order corresponds to increasing the angle). The operating current of the slave laser and the seed power are $204\ \mathrm{mA}$ and $5\ \mathrm{mW}$, respectively.  The same dependence is observed for the horizontal angle of the Mirror 2. }
 \label{fig2}
 \end{center}
\end{figure}

Figure \ref{fig2} shows the variation of the cavity-transmission signals as the vertical angle of the Mirror 2 is gradually changed at an operating current of 204 mA. Note that the increasing alphabetical order in the figures corresponds to increasing the vertical angle of the Mirror 2. One might think that the injection locking is successfully achieved under the condition corresponding to Fig. \ref{fig2} (b) or (f), because  a single peak is monitored within one FSR. However, the heights of the signal peaks are approximately 30$\%$ of that when the same intensity of the seed light is fed into the Fabry-Perot cavity, which means 70$\%$ of the slave laser output still keeps broad spectrum. Note that no peaks are monitored when the angle of the mirror is changed too much [see Fig. \ref{fig2} (a) and (g)], since the spatial mode matching between the seed and the slave lights is lost.

\begin{figure}[htbp]
 \begin{center}
  \includegraphics{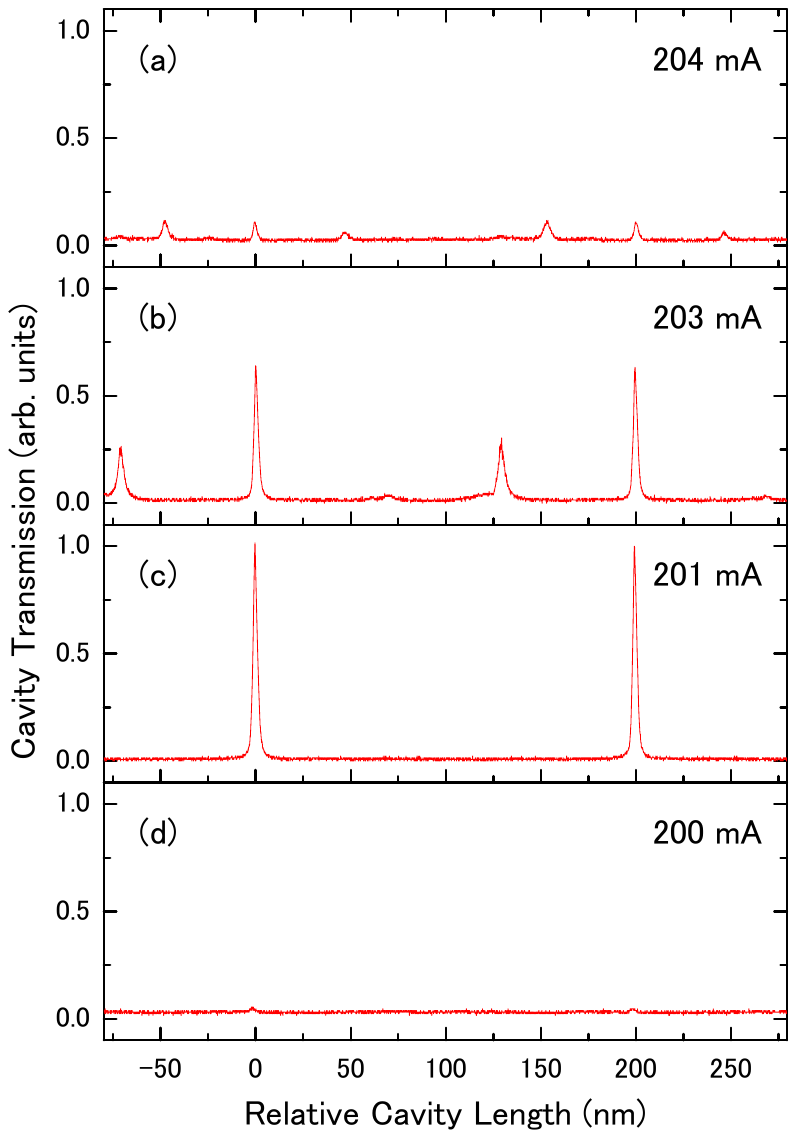}
 \caption{Dependence of the cavity-transmission signals on the injection current for the slave laser. The injection currents are (a) $204\ \mathrm{mA}$, (b) $203\ \mathrm{mA}$, (c) $201\ \mathrm{mA}$ and (d) $200\ \mathrm{mA}$, respectively.}
 \label{fig3}
 \end{center}
\end{figure}

The dependence of the cavity transmission signals on the inclination of the Mirror 2 is symmetric with respect to the specific angle [Fig. \ref{fig2} (d)]. The same behavior is observed when the horizontal angle of the Mirror 2 is changed, i.e., the dependence of the signals is also symmetric with respect to the specific horizontal angle. We find out that the injection locking can be optimized by slightly decreasing the slave laser current after adjusting both the vertical and horizontal angles of the Mirror 2 to these specific angles. Figure \ref{fig3} shows the dependence of the cavity-transmission signals on the injection current for the slave laser under this condition. The height of the signal peaks increases as the injection current decreases and coincides with that when the same intensity of the seed light is fed into the cavity [Fig. \ref{fig3} (c)], i.e., the single-mode emission is obtained for the slave laser at the injection current of $201\ \mathrm{mA}$. Even if the current is further decreased, the single-mode emission is maintained for an interval of $1\ \mathrm{mA}$, and consequently, additional techniques to stabilize the current are not required. When the current is decreased too much, the injection locking is lost and no peaks are monitored [see Fig. \ref{fig3} (d)].

\begin{figure}[htbp]
 \begin{center}
  \includegraphics{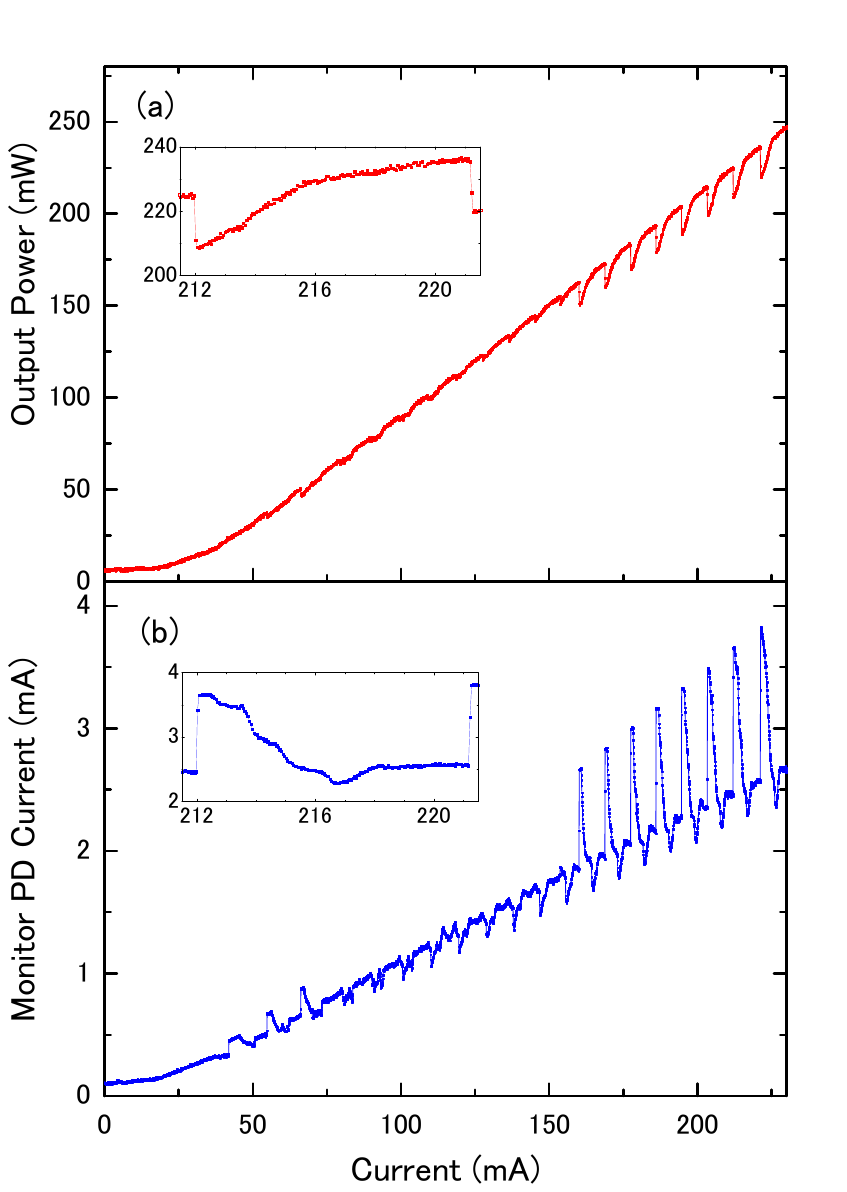}
 \caption{(a) Slave laser output power and (b) monitor photo-diode current when the injection current is decreased under the condition of injection locking. Insets: enlarged graphs situated between $211\ \mathrm{mA}$ and $222\ \mathrm{mA}$.}
 \label{fig4}
 \end{center}
\end{figure}

Figure \ref{fig4} shows the variation of the output power of the slave laser when the injection current is decreased. As the injection current decreases within a range between $212.0\ \mathrm{mA}$ and $221.2\ \mathrm{mA}$, the monitor photodiode current (which is proportional to the back-end power of the slave laser) increases  [Fig. \ref{fig4} (b)], whereas the slave output power decreases [Fig. \ref{fig4} (a)]. The condition of the injection locking is optimized by decreasing the injection current, which results in negative interference between the slave output and the seed light reflected from the front facet of the slave. On the other hand, positive interference occurs between the seed and the slave lights at the back-end of the slave chip, which leads to an increment of the monitor photodiode current. As can be seen in Fig. \ref{fig4}, such interferences are repeated with a cycle of approximately $9\ \mathrm{mA}$.
\section{Frequency scanning of the injection locked slave laser}
\label{sec4}
\begin{figure*}[htbp]
 \begin{center}
  \includegraphics
 {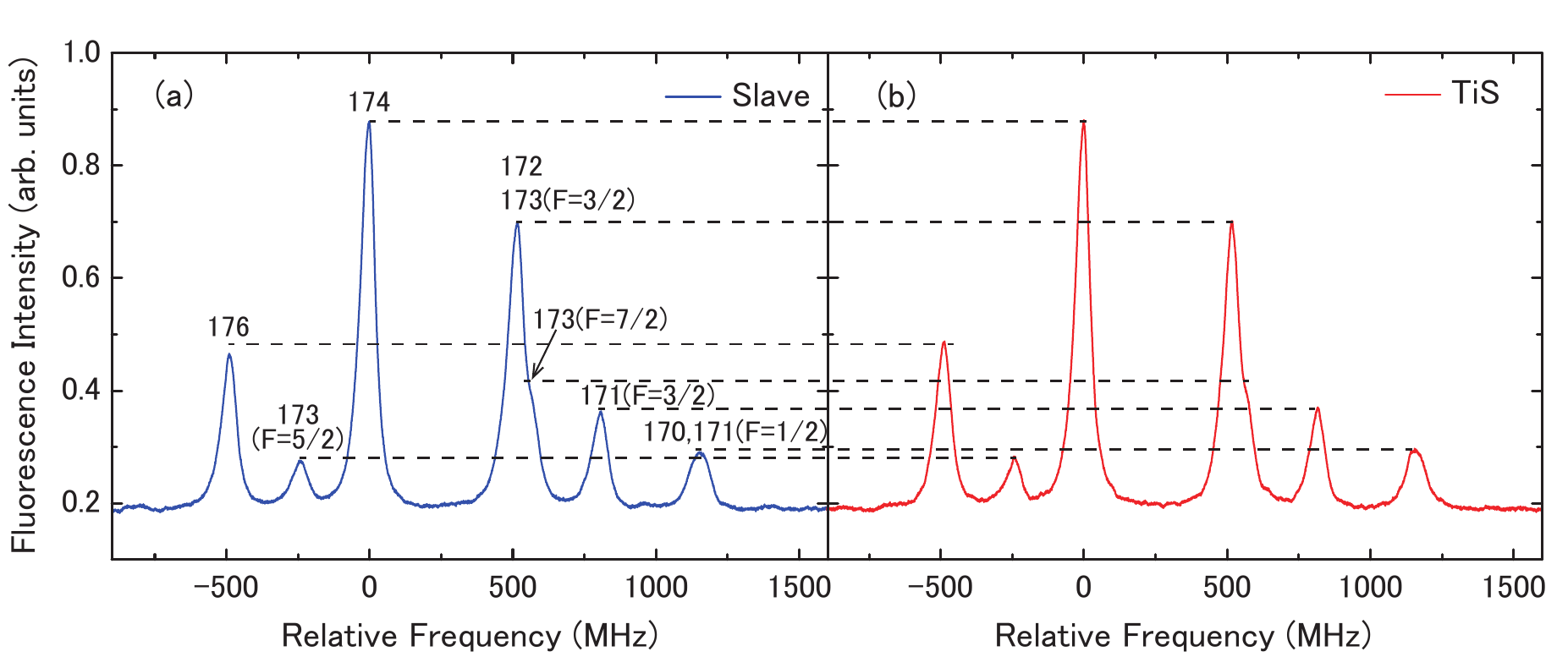}
 \caption{Fluorescence from the Yb atomic beam irradiated by (a) the slave laser and (b) the frequency doubled Ti:Sapphire laser as a function of the laser frequency. }
 \label{fig5}
 \end{center}
\end{figure*}

In order to confirm that the slave laser follows up the seed light along a wide range of frequency, fluorescence from the Yb atomic beam is monitored while sweeping the frequency of the seed light. To realize a wide range of frequency scanning without mode hopping, here a frequency doubled Ti:Sapphire laser is employed as the seed light. The atomic beam is irradiated in turns, first by the seed light and later by the injection locked slave laser, in a direction perpendicular to that of the atomic beam. The laser intensity is regulated to $2.0\ \mathrm{mW/cm^2}$ and the corresponding saturation parameter $3.5 \times 10^{-2}$ is well below the unity.  If the injected slave laser is successfully emitting in single frequency mode, the fluorescence intensity is proportional to only its power. Figure \ref{fig5} shows the obtained fluorescence intensity, where the signal heights coincide in both cases. The result confirms that the slave laser follows up the seed light keeping single-mode emission within the scan range of $1.5\ \mathrm{GHz}$ covering whole spectra corresponding to the stable isotopes of Yb.
\section{Conclusion}
\label{sec5}
In summary, we developed an laser diode based high power $399\ \mathrm{nm}$ light source for possible applications on laser cooling of Yb atoms, where a final output of $220\ \mathrm{mW}$ was obtained through injection locking technique. The injection locking condition is systematically optimized by using the signal of a scanning Fabry-Perot cavity. Since only $5\ \mathrm{mW}$ of seed light is required for implementing the injection locking, multiple high-power light sources can be prepared by using a single master laser. Our method is thus useful for implementing transportable optical lattice clocks, and investigating condensed matter physics or quantum information processing using cold atoms.
\section{Acknowledgements}
\label{sec6}
The authors thank Naoki Tambo and Yuki Miyazawa for technical assistance. This work is supported by a Grant-in-Aid for Scientific Research on Innovative Areas “Fluctuation $\&$ Structure” from the MEXT, the Cabinet Office, Government of Japan, through its “Funding Program for Next Generation World-Leading Researchers”, the Matsuo Foundation and the Murata Science Foundation. One of the authors (M.M.) was supported in part by the Japan Society for the Promotion of Science.
%

% %\section*{References}

\bibliography{mybibfile}

\end{document}